\documentclass{emulateapj}

\newcommand{\expnt}[2]{\ensuremath{#1 \times 10^{#2}}}   
\newcommand{\gsim}{\gtrsim}
\newcommand{\lsim}{\lesssim}

\newcommand{\xmm}{\textit{XMM}}
\newcommand{\hr}{\ensuremath{^{\rm h}}}
\newcommand{\mn}{\ensuremath{^{\rm m}}}

\newcommand{\rxj}{RX~J0720.4\ensuremath{-}3125}
\newcommand{\cxo}{\textit{CXO}}
\newcommand{\chandra}{\textit{Chandra}}
\newcommand{\rosat}{\textit{ROSAT}}

\newcommand{\secsec}{\ensuremath{\mbox{s s}^{-1}}}
\newcommand{\Hzsec}{\ensuremath{\mbox{Hz s}^{-1}}}

\defcitealias{kkvkm02}{Paper I}

\newcommand{\fdot}{\ensuremath{\dot \nu}}
\newcommand{\fdotdot}{\ensuremath{\ddot \nu}}

\slugcomment{To appear in ApJL}
\shorttitle{A Timing Solution for RX~J0720.4$-$3125}
\shortauthors{Kaplan \&\ van~Kerkwijk}
\begin{document}

\title{A Coherent Timing Solution for the Nearby Isolated Neutron Star
RX~J0720.4$-$3125}
\author{D.~L.~Kaplan\altaffilmark{1,2} and M.~H.~van Kerkwijk\altaffilmark{3}}

\altaffiltext{1}{Pappalardo Fellow}
\altaffiltext{2}{Center for Space
Research, Massachusetts Institute of Technology, 77 Massachusetts Avenue,
37-664D, Cambridge, MA 02139, USA; dlk@space.mit.edu}
\altaffiltext{3}{Department of Astronomy and Astrophysics, University
  of Toronto, 60 St.\ George Street, Toronto, ON M5S 3H8, Canada;
mhvk@astro.utoronto.ca}

\begin{abstract}
We present the results of a dedicated effort to measure the spin-down
rate of the nearby isolated neutron star \rxj.  Comparing arrival
times of the 8.39-sec pulsations for data from \chandra\ we
derive an unambiguous timing solution for \rxj\ that is accurate to
$<0.1$~cycles over $>5$~years.  Adding data from \xmm\ and \rosat, the final
solution yields $\dot P=\expnt{(6.98\pm0.02)}{-14}\,\secsec$; for
dipole spin-down, this implies a characteristic age of 2~Myr and a
magnetic field strength of $\expnt{2.4}{13}$~G.  The phase residuals
are somewhat larger than those for purely regular spin-down, but do
not show conclusive evidence for higher-order terms or a glitch.  From
our timing solution as well as recent X-ray spectroscopy, we concur
with recent suggestions that \rxj\ is most likely an off-beam radio
pulsar with a moderately high magnetic field.
\end{abstract}
\keywords{pulsars: individual (RX J0720.4$-$3125)
      --- stars: neutron
      --- X-rays: stars}

\section{Introduction}

One of the interesting results from \rosat\ All-Sky Survey \citep{rbs}
was the discovery of seven objects that appear to be nearby,
thermally-emitting neutron stars that have little if any
magnetospheric emission (see \citealt{haberl04} for a review).  These
objects, known most commonly as ``isolated neutron stars,'' are
distinguished  by their long periods ($\gsim 3$~s,
when measured), largely thermal spectra with cool temperatures ($kT
\lsim 100$~eV), faint optical counterparts (when detected), and lack
of radio emission.

It is not yet clear what sets the isolated neutron stars apart from
the nearby, relatively young rotation-powered pulsars that also have
cool thermal emission --- sources like PSR~B0656+14 and
PSR~B1055$-$52--- which tend to have short ($<1$~s) spin periods,
$\sim 10^{12}$-G magnetic fields, non-thermal (i.e.\ power-law)
components in their X-ray
spectra, and radio pulsations \citep[e.g.,][]{pz03,kaplan04}.  The isolated
neutron stars are known to have longer periods, but their spin-down
rates (and hence magnetic fields ) are unknown, largely because it has
not yet been possible to determine a reliable, coherent timing
solution (\citealt{kkvkm02}, hereafter \citetalias{kkvkm02};
\citealt{zhc+02}).  

In this \textit{Letter} we report a new analysis of the variations of
the 8.39-s period of the second brightest source of the group, \rxj\
\citep{hmb+97}.  We describe our analysis of \chandra\ data obtained
specifically for timing purposes, as well as archival \rosat,
\chandra, and \xmm\ data, in \S~\ref{sec:obs}.  In
\S~\ref{sec:timing}, we show that with the new data, we can avoid the
pitfalls of the previous phase-coherent timing analyses and obtain a
reliable timing solution.  We discuss possible timing noise in
\S~\ref{sec:noise} and the implications of our result in
\S~\ref{sec:discuss}.
\ \\

\section{Observations}\label{sec:obs}

Our primary data were eight observations with the Advanced CCD Imaging
Spectrometer \citep[ACIS;][]{gbf+03} aboard the \textit{Chandra X-ray
Observatory} \citep[\cxo;][]{wtvso00}.  These were 
designed for timing accuracy, consisting of two sets of four exposures
geometrically spaced over a period of about two weeks and separated
by about half a year.  We combined these with data from other
\chandra\ observations, as well as from observations with
\textit{XMM-Newton} \citep{jla+01} and \rosat\ \citep{trumper93}.  A
log of all observations is given in Table~\ref{tab:obs}.

For the \chandra\ data, we processed the level-1 event lists to the
level-2 stage following standard procedures and the latest calibration
set (CALDB version 3.0.0).  For the ACIS continuous-clocking data,
this includes correcting the recorded event times for readout, dither,
and spacecraft motion --- corrections that used to require additional
steps \citep{zpst00}.  We extracted events within $1\arcsec$ of the
source, and then applied a clock correction of $284.7\,\,\mu{\rm s}$
to the arrival times \citep*{dhm03}; the arrival times should now be
accurate to $\lsim 6\,\,\mu{\rm s}$.  For the HRC-S/LETG data, we
extracted zeroth-order events from a circle with radius 10 pixels
($1\farcs3$), and first-order events using the standard LETG spectral
extraction windows, but limited to $10\leq\lambda\leq65\mbox{\AA}$.
Finally, we used the \texttt{axbary} program to barycenter all of the
events (using the optical position: $\alpha_{\rm
  J2000}=07\hr20\mn24\fs96$, $\delta_{\rm
  J2000}=-31\degr25\arcmin50\farcs2$; \citealt{kvkm+03})

\begin{deluxetable*}{lrcrrc}[t]
\tablewidth{0pt}
\tablecaption{Log of Observations and Times of Arrival\label{tab:obs}}
\tablehead{
\colhead{Instrument\tablenotemark{a}}&
\colhead{ID\tablenotemark{b}}&
\colhead{Date}&
\colhead{Exp.}&
\colhead{Counts}&
\colhead{TOA\tablenotemark{c}}\\
&&&\colhead{(ks)}&& \colhead{(MJD)}\\[-2.2ex]
}
\startdata
PSPC\dotfill       & 338   & 1993-09-27 &   3.2 &  5800&49257.2547031(25)\\
HRI\dotfill        & 884   & 1996-11-03 &  33.7 & 12662&50391.3006530(16)\\
HRI\dotfill        & 944   & 1998-04-20 &   8.1 &  3074&50925.6881172(36)\\
HRC\dotfill        & 368   & 2000-02-01 &   5.4 &  3472&51575.3026910(46)\\
HRC\dotfill        & 745   & 2000-02-02 &  26.1 & 15149&51576.2804856(27)\\
HRC\dotfill        & 369   & 2000-02-04 &   6.1 &  3667&51578.7722735(65)\\
PN/ff/thin\dotfill &  78-S3& 2000-05-13 &  21.1 &144104&51677.2260789(\phn5)\\
MOS2/thin\dotfill  &  78-S2& 2000-05-13 &  43.9 & 73915&51677.4127431(\phn7)\\
PN/ff/med\dotfill  & 175-S3& 2000-11-21 &  25.7 &153037&51869.8413358(14)\\
MOS1/open\dotfill  & 175-U2& 2000-11-21 &   6.8 & 17762&51869.8433759(14)\\
MOS2/open\dotfill  & 175-U2& 2000-11-21 &   7.2 & 21084&51869.9571032(\phn6)\\
ACIS-CC\dotfill    &2774   & 2001-12-04 &  15.0 & 31831&52247.7881789(11)\\
ACIS-CC\dotfill    &2773   & 2001-12-05 &  10.6 & 22847&52248.2835843(13)\\
ACIS-CC\dotfill    &2772   & 2001-12-06 &   4.1 &  8790&52249.6286894(26)\\
PN/ff/thin\dotfill & 533-S3& 2002-11-06 &  28.3 &199841&52584.9260561(\phn5)\\
PN/ff/thin\dotfill & 534-S3& 2002-11-08 &  30.2 &212177&52587.0013053(\phn4)\\
MOS1/open\dotfill  & 622-U2& 2003-05-02 &   7.6 & 17629&52761.6222174(14)\\
MOS2/open\dotfill  & 622-U2& 2003-05-02 &   7.5 & 18788&52761.6226056(12)\\
PN/sw/thick\dotfill& 622-U2& 2003-05-02 &  72.8 &210160&52761.9950589(\phn5)\\
PN/sw/thin\dotfill & 711-S7& 2003-10-27 &  18.1 &112876&52939.8228751(\phn9)\\
PN/sw/thick\dotfill& 711-S8& 2003-10-27 &  25.0 &138689&52939.8228774(\phn8)\\
MOS1/open\dotfill  & 711-U2& 2003-10-27 &  13.8 & 33323&52939.8506513(\phn5)\\
MOS2/open\dotfill  & 711-U2& 2003-10-27 &  13.8 & 35636&52940.1162720(\phn5)\\
ACIS-CC\dotfill    &4666   & 2004-01-06 &  10.1 & 19048&53010.2635608(14)\\
ACIS-CC\dotfill    &4667   & 2004-01-07 &   4.8 &  8938&53011.2639869(20)\\
ACIS-CC\dotfill    &4668   & 2004-01-11 &   5.2 &  9334&53015.5407400(19)\\
ACIS-CC\dotfill    &4669   & 2004-01-19 &   5.2 &  9391&53023.1274147(23)\\
HRC\dotfill        &5305   & 2004-02-27 &  35.7 & 21597&53062.4142490(27)\\
PN/ff/thin\dotfill & 815-S1& 2004-05-22 &  31.6 &219855&53147.6811948(\phn4)\\
ACIS-CC\dotfill    &4670   & 2004-08-03 &  10.1 & 17432&53220.9975987(14)\\
ACIS-CC\dotfill    &4671   & 2004-08-05 &   5.1 &  8051&53222.2171299(21)\\
ACIS-CC\dotfill    &4672   & 2004-08-09 &   5.1 &  8556&53226.2443808(25)\\
ACIS-CC\dotfill    &4673   & 2004-08-23 &   5.1 &  7133&53240.1824669(24)\\
HRC\dotfill        &5581   & 2005-01-23 &  67.7 & 44801&53393.6657119(18)\\
\enddata
\tablenotetext{a}{PSPC: Position Sensitive Proportional Counter
    \citep{bp95} aboard \rosat.  
  HRI: High-Resolution Imager \citep{zdhk95} aboard \rosat.
  HRC: High-Resolution Camera for spectroscopy aboard \chandra\ (HRC-S;
    \citealt{kck+97}), used with the Low-Energy Transmission Grating.
  ACIS: \chandra's Advanced CCD Imaging Spectrometer \citep{gbf+03},
    used in continuous clocking mode.
  EPIC-pn: \xmm's European Photon Imaging Camera with PN detectors
    \citep{sbd+01}, used in full-frame (ff) or small window (sw)
    mode, with thin, medium, or thick filter.
  EPIC-MOS1/2: European Photon Imaging Cameras with MOS detectors
    aboard \xmm\ \citep{taa+01}, used in small-window mode with thin
    or no (open) filter.} 
\tablenotetext{b}{Observation identifier (\cxo, \rosat) or revolution
  number and exposure identifier (\xmm).} 
\tablenotetext{c}{The TOA is defined as the time of maximum light
  closest to the middle of each observation, and is given with
  1-$\sigma$ uncertainties.} 
\end{deluxetable*}

For the \xmm\ data, we used the standard procedures \texttt{emchain}
and \texttt{epchain} (XMMSAS version 6.1.0) to reprocess the
observations.  One additional step was necessary for the PN data set
622-U2, for which we found a small number of duplicate events (frames
963685--963719); we removed these before processing.  Next, we
extracted all single-pixel events within 37\farcs5 of the source
position, and used \texttt{barycen} to convert the arrival times to
the solar-system barycenter\footnote{Some portions of the 2000 and
  2002 \xmm/PN observations were affected by a known processing
  problem that rejected significant portions of the observations; see
  \url{http://xmm.vilspa.esa.es/sas/documentation/watchout/lost\_events.shtml}.
  This should not introduce any systematic error, though it means that
  our TOAs for these observations are not as precise as possible with
  all events.  However, since the present TOA uncertainties are smaller than
  the timing noise (\S~\ref{sec:noise}), we decided not to
  try to remedy this problem.}.

The reduction of the \rosat\ data followed that in
\citetalias{kkvkm02}, except that we properly corrected the event
times to the Barycentric Dynamical Time (TDB) system instead of the
Coordinated Universal Time (UTC) system returned by the
\texttt{FTOOLS} barycentering tasks (see \citealt{chzz04}).  We
used the corrections supplied in \citet[][p.\ 14]{allen}.

\ \\
\section{Timing Analysis}
\label{sec:timing}
Our goal was to use times-of-arrival (TOAs) to infer a phase-coherent
timing solution involving the spin period and its derivative, where
each cycle of the source was counted.  To measure TOAs we needed an
initial reference period, something which we determined using a
$Z_1^2$ test \citep[Rayleigh statistic;][]{bbb+83} on the combined
2004 January ACIS and 2004 February HRC data.  We find
$P=8.3911159(10)$~s (here and below, numbers in parentheses indicate
twice the formal 1-$\sigma$ uncertainties in the last digit unless otherwise
indicated), which is consistent with our earlier value
(\citetalias{kkvkm02}) but much more accurate because we could
coherently connect observations  over a much longer (52~day)
time span.

Using this period, we constructed binned light curves (with 16 phase
bins) and determined the TOAs by fitting a sinusoid (appropriate for
the sinusoidal pulsations of \rxj; \citealt{hmb+97}); the
uncertainties were calculated from the uncertainties in the phases of
the fitted sinusoids (we verified that we obtain consistent results if
we change the binning or measure TOAs using cross-correlation
instead).  Here, the TOA is defined as the time of maximum light
closest to the middle of the observation, a choice which minimizes
co-variance with small changes in period.  We present TOAs for all of
the data in Table~\ref{tab:obs}.

We then determined a timing solution for only the \chandra\ data
using an iterative procedure.  We first used the reference period to
determine cycle counts for the five above-mentioned observations, as
well as the next observation closest in time.  We fit these cycle
counts to
\begin{equation}
\phi(t) = \phi_0 + \nu (t-t_0) + \frac{1}{2} {\dot \nu}(t-t_0)^2 + \frac{1}{6}{\ddot \nu}(t-t_0)^3\ldots,
\label{eqn:phi}
\end{equation}
where $\phi_0$ is the cycle count plus phase at reference time $t_0$,
$\nu$ is the spin-frequency, $\dot \nu$ is its derivative, $\ddot \nu$
is the second derivative.  We then
iterated, using the improved solution to determine the cycle count for
the next observation, etc.  We started with $\fdot=\fdotdot=0$, but
left $\fdot$ free once that significantly improved the fit; $\fdotdot$
was not required (cf., \S~\ref{sec:noise} below).

The final ephemeris listed in Table~\ref{tab:ephem} has small,  
$\leq0.1$~cycle, residuals, and fits the \chandra\ data well:
$\chi^2_\nu\equiv\chi^2/N_{\rm dof}=1.06$ (with $N_{\rm dof}=13$
degrees of freedom).  To test the uniqueness of our ephemeris, we
tried changing the cycle counts (adding or subtracting one or more
cycles) at the least unambiguous points, but
found that the resulting solutions were very poor (e.g., altering the
cycles between the 2001 ACIS and 2000 HRC observations gave
$\chi^2_{\nu}=89.26$).

\begin{figure}
\plotone{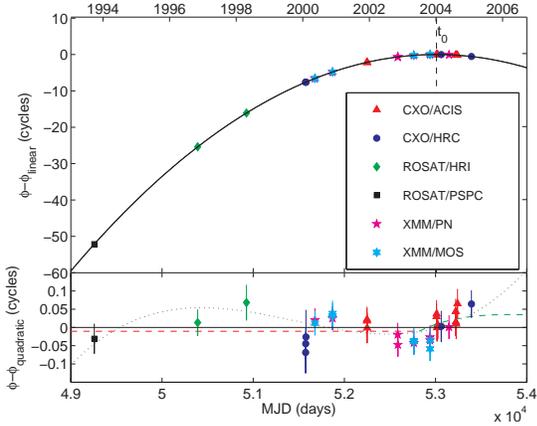}
\caption{Phase residuals for \rxj.  The top panel shows the residuals
  for each TOA compared to a linear ($\fdot=0$) model.  The solid
  curve gives the best-fit quadratic ($\fdot \neq 0$, $\fdotdot=0$)
  ephemeris for all data (Tab.~\ref{tab:ephem}, with a 0.27-s
  systematic uncertainty added in quadrature).  The vertical dashed
  line indicates the reference time $t_0$.  The bottom panel shows the
  residuals relative to the quadratic model.  We also show a best-fit
  cubic model ($\fdotdot \neq 0$; dotted line) and a model that
  includes a glitch in frequency near MJD~52800 ($\Delta f\approx
  \expnt{1}{-9}$; dashed curve).}
\label{fig:resid}
\end{figure}

To improve and extend the ephemeris, we added the \xmm\ and \rosat\
data into the solution (see Table~\ref{tab:ephem}).  As for the
\chandra\ data alone, the cycle counts are unambiguous and, as can be
seen in Fig.~\ref{fig:resid}, the residuals remain below $0.1$~cycles,
lending additional credence to our results.  For verification, we also
examined the fit using just the \chandra\ and \xmm\ data, which avoids
the large gaps between the \rosat\ observations.  We found the same
cycle counts, but a statistically significant difference in the
inferred parameters [$\nu=0.11917366887(12)$~Hz and
$\fdot=-\expnt{9.69(4)}{-16}\,\,\Hzsec$, while the fit to all data
gave $\nu=0.1191736700(2)$ and $\fdot=-\expnt{9.915(14)}{-16}\,\,\Hzsec$].

While we are confident in our fit in general, the above discrepancy is
puzzling.  Furthermore, the $\chi^2_\nu$ values for the fits including
the \xmm\ data are poor, with $\chi^2_\nu=6.16$ for \chandra$+$\xmm\
($N_{\rm dof}=28$; ${\rm rms}=0.34$~s) and $\chi^2_{\nu}=10.25$ for
all of the data ($N_{\rm dof}=31$; ${\rm rms}=0.36$~s), while the
solution for \chandra$+$\rosat, although somewhat ambiguous, is
tolerable ($\chi^2_\nu=1.40$) and similar to the \chandra-only
solution.  The $\chi^2$ values for the full fits are dominated by the
contribution of the \xmm\ data, which have very small formal
uncertainties (but are not always entirely consistent from one
instrument to another for the same observation; see
Tab.~\ref{tab:obs}).  The \xmm\ data also cause the values of $\fdot$
to differ by 3\% and the \chandra$+$\xmm\ ephemeris to be a poor match
to the \rosat\ points.  We discuss these deviations in more detail
below; here, we note that for the estimates of the values and
uncertainties in Table~\ref{tab:ephem}, we ensured $\chi^2_\nu\simeq1$
by adding in quadrature an additional uncertainty of 0.27~s to all
TOAs.  We stress, however, that our overall solution is robust, and
the inferred spin-down rate should be reliable at the $<\!3$\% level.

\begin{deluxetable}{l c c}
\tablewidth{0pt}
\tablecaption{X-ray Timing Ephemerides for \rxj\label{tab:ephem}}
\tablehead{
\colhead{Quantity\tablenotemark{a}} & \colhead{\cxo\ only} & 
\colhead{All data\tablenotemark{b}}\\[-2.2ex]
}
\startdata
Dates (MJD) \dotfill         & 51575--53394 & 49257--53394\\
$t_{0}$ (MJD)\dotfill        & 53010.2635605(18) & 53010.2635637(24)\\
$\nu$ (Hz) \dotfill          & 0.11917366926(46)  & 0.11917366908(38)\\
$\dot \nu$ (\Hzsec) \dotfill & $\expnt{-9.97(11)}{-16}$ & $-\expnt{9.918(30)}{-16}$\\
TOA rms (s) \dotfill         & 0.18 & 0.31\\
$\chi^2$/DOF \dotfill        & 13.8/13=1.06 & 30.7/31=0.99\\[1ex]
$P$ (s)\dotfill              & 8.391115305(32) & 8.391115532(26)\\
$\dot P$ (\secsec)\dotfill   & $\expnt{7.019(80)}{-14}$ & $\expnt{6.983(22)}{-14}$\\
$\dot E$ ($\mbox{erg s}^{-1}$)\dotfill& $\expnt{4.7}{30}$ & $\expnt{4.7}{30}$\\
$B_{\rm dip}$ (G) \dotfill   & $\expnt{2.5}{13}$ & $\expnt{2.4}{13}$\\
$\tau_{\rm char}$ (yr)\dotfill& $\expnt{1.9}{6}$ & $\expnt{1.9}{6}$\\
\enddata
\tablecomments{Uncertainties quoted are twice the formal 1-$\sigma$
  uncertainties in the fit.}
\tablenotetext{a}{$\dot E=10^{45}I_{45}4\pi^2\nu\dot\nu$ is the 
  spin-down luminosity (with $I=10^{45}I_{45}\,\,{\rm g}\,\,{\rm
    cm}^{2}$ the moment of 
  inertia); $B_{\rm dip}=\expnt{3.2}{19}\sqrt{P{\dot P}}$ is the
  magnetic field inferred assuming spin-down by dipole radiation;
  $\tau_{\rm char}=P/2{\dot P}$ is the characteristic age, 
  assuming an initial spin period $P_0\ll P_{\rm now}$.}
\tablenotetext{b}{A 0.27-s systematic uncertainty has been added in
  quadrature to these data to account for timing noise.  See
  \S~\ref{sec:noise}.} 
\end{deluxetable}

\section{Deviations from Regular Spin Down?}
\label{sec:noise}
There could be several reasons for the relatively poor fit of the
timing solutions to our full set of arrival times.  First, \rxj\ may
show some rotational instabilities or ``timing noise,'' as seen in
radio pulsars.  We can estimate the magnitude of the long-term timing
noise by fitting the phase residuals with a third-order
($\fdotdot\neq0$) solution like that shown in Fig.~\ref{fig:resid}.
This does in fact improve the fit --- giving $\chi^2_\nu=6.02$ for
$N_{\rm dof}=30$ with an rms of 0.32~s --- though it is still formally
unacceptable.  We find $\fdotdot\approx \expnt{2.4(4)}{-25}\mbox{ Hz
  s}^{-2}$ (this also changes $\fdot$ by 3\% from the value in
Tab.~\ref{tab:ephem}), giving a timing noise measurement of
$\Delta_8\equiv \log_{10}(|{\ddot \nu}|(10^8)^3/6\nu)= -0.5$.  This
value is on the high side of, but not outside the range expected from
relations between $\dot P$ and $\Delta_8$ found for radio pulsars
\citep{antt94} and magnetars \citep{wkf+00,gk02}.  We do not believe
it is likely that the third-order solution represents a real long-term
change in $\fdot$, since the value of $\fdotdot$ changes significantly
if one uses just the \chandra, \chandra$+$\xmm, or all of the
data\footnote{From Fig.~\ref{fig:resid}, it may appear that one could
  obtain a better fit by reducing the cycle count for the 1998 HRI
  point by one; however, doing that, the other \rosat\ points cannot be
  reproduced any more.}  [8(6), 18(4), or $\expnt{2.4(4)}{-25}$
  $\mbox{Hz s}^{-2}$, with $\chi^2_\nu=0.64$, $2.81$, or $6.02$
  respectively].

A second possibility is a sudden change in rotation --- a glitch, as
proposed (and largely rejected) by \citet{dvvmv04} to account for
variations in spectral shape (differences of 10\% in the inferred
temperature) and pulse shape (changes of 50--100\% in pulsed fraction)
observed between 2000 and 2003.  The systematic increase in the
residuals after MJD~53000 in Figure~\ref{fig:resid} might indeed
indicate a glitch.  We tried fitting a simple glitch model, in which
we assumed that only the frequency changed, that the recovery time was
longer than the span of our observations, and that the glitch occurred
on 2003 July 1 (MJD~52821), in between the two 
\xmm\ observations that showed the largest spectral change
\citep{dvvmv04}.  We find a reasonable fit for a glitch with $\Delta
f=\expnt{1.3}{-9}$~Hz and a recovery time $>3$~yr, giving
$\chi^2_\nu=7.7$ (see Fig.~\ref{fig:resid}).  This would be a small
glitch, with $\Delta f/f=\expnt{1.1}{-8}$ compared to values of
$10^{-9}$ to $10^{-6}$ for radio pulsars \citep*[with smaller values
  more typical for larger magnetic fields;][]{lss00}.  It also implies
that energetically, the putative glitch would be insignificant: the
change in kinetic energy of $\sim 10^{36}$~ergs \citep{vrem91} would
only be noticeable if dissipated in $<1$~day given the
bolometric luminosity of \rxj\ of $\expnt{2}{32}d_{300}^2\mbox{ ergs
  s}^{-1}$ (where $d=300d_{300}$~pc is the distance to \rxj;
\citealt{kvkm+03}).  In principle, however, it could still have
altered the light curve and spectrum of \rxj\ through realignment of
the magnetic field relative to the spin axis.

Finally, a more mundane explanation for the relatively poor fit is
that the data are from different instruments with different energy
responses (even among a single instrument aboard \xmm, the changing
filters alter the response).  The pulse profile of \rxj\ is known to
depend on energy (\citealt{czr+01}; \citetalias{kkvkm02};
\citealt{hztb04}) and to change over time \citep{dvvmv04}.  While the
changes to the shape are small, some systematic offsets are expected between the
pulse profiles as measured by different instruments or at different
times.  We hope to investigate this in more detail in the near
future.  

At present, we cannot distinguish between the various possibilities
for the relatively poor fits.  The predicted future behavior is
different, however, and thus further observations of \rxj\ (some of
which are in progress) should be able to distinguish between these
models.

\section{Discussion \& Conclusions}
\label{sec:discuss}

From our timing solution, we infer a spin-down rate $\dot
P=6.98(2)\times10^{-14}~\secsec$.  This is consistent with the
limits derived in \citetalias{kkvkm02} and by \citet{zhc+02}, but
inconsistent with the tentative solution of \citet{chzz04}, who found
$\dot P=\expnt{(1.4\pm0.6)}{-13}\,\,\secsec$ at 99\% confidence (but
who noted that elements of their solution were inconsistent with each
other and that their analysis was subject to confusing aliases).  In
Table~\ref{tab:ephem}, we list derived parameters --- rotational
energy loss rate, magnetic field, and characteristic age (the latter
two under the assumption of magnetic dipole spin-down).

The values of $P$ and $\dot P$ place \rxj\ well above above most of
the radio-pulsar ``death-lines'' proposed so far
\citep*[e.g.,][]{ymj99} and in a region populated by radio pulsars in
$P$-$\dot P$ diagrams like that in \citetalias{kkvkm02} (its
parameters are approximately between those of PSR~J1830$-$1135, with
$P=6.2$~s, $\dot P=\expnt{5}{-14}\,\secsec$, and PSR~J1847$-$0130,
with $P=6.7$~s, $\dot P=\expnt{1.3}{-12}\,\secsec$).  Hence, \rxj\ may
well be a radio pulsar itself, but one whose radio beam(s) do not
intersect our line of sight.  Its inferred magnetic field,
$B=\expnt{2.4}{13}$~G, is not exceptional; the Parkes Multi-beam
Survey \citep{mlc+01} in particular has discovered a fair number of
radio pulsars with $B\gsim 10^{13}$~G
\citep[e.g.,][]{ckl+00,mhl+02,msk+03}, and it is now clear the
distribution of magnetic fields is flatter than previously assumed
\citep{vml+04}.

With $\dot E=\expnt{4.7}{30}\mbox{ ergs s}^{-1}$, \rxj\ is not
expected to have much non-thermal X-ray emission: from the relation of
\citet{bt97}, one estimates $L_{\rm X,\,non-th}\sim 10^{-3}\dot E
\simeq \expnt{5}{27}\mbox{ ergs s}^{-1}$, much smaller than the
thermal emission, $L_{\rm X,\,therm}\simeq \expnt{2}{32}d_{300}\mbox{
ergs s}^{-1}$.  This is consistent with limits from \chandra\ and
\xmm\ (\citealt{pmm+01}; \citealt*{pzs02}; \citealt{kvkm+03}).

What is somewhat puzzling is the inferred age of 2~Myr.  Tracing \rxj\
back to OB associations where it might have been born put it close to
the Trumpler~10 association $\sim\!0.7$~Myr ago (\citealt*{mzh03};
\citealt{kaplan04}).  Similarly, based on its estimated temperature
and luminosity, most standard cooling models (modified URCA for
$1.4~M_{\odot}$ neutron stars) put \rxj\ at $\lsim1$~Myr
(\citealt{hh98}; \citetalias{kkvkm02}; \citealt{chzz04}).

It is of course possible that \rxj\ was born with a long period and/or
had significantly non-dipole braking, such that the spin-down age is
not a good estimate of its true age.  However, no case with as long a
birth period as would be required for \rxj\ is known among radio
pulsars \citep[cf.][]{klh+03,gkr04}.

Another possible explanation is that \rxj\ was ejected from a binary
system with a massive companion $\sim 0.7$~Myr ago, either when the
companion underwent a supernova or during a binary exchange interaction.
In this case, a relatively long period is expected: if the neutron
star accreted matter from its companion, its spin period would have
tended toward the equilibrium period, $P_{\rm{}eq}\approx
5{\rm\,s}\,(B/10^{13}{\rm\,G})^{6/7}(\dot M/\dot M_{\rm Edd})^{-3/7}$
(where $\dot M$ is the accretion rate and $\dot M_{\rm Edd}$ is the
Eddington rate).  A relatively short cooling age would also be
consistent with this model: the accretion and accompanying steady
hydrogen burning could reheat the neutron star (or keep it hot).  Of
course, it remains to be seen that a suitable evolutionary scenario
can be found.  In any case, the prediction for the model with two
supernovae is that there may well be another $\la\!1~$Myr old neutron
star whose proper motion traces back to the same origin as \rxj\
\citep[cf.][]{vcc04}.

Finally, we can compare the magnetic field strength of
$\expnt{2.4}{13}$~G with what is inferred from the broad absorption
feature in the spectrum (observed to be at 0.3~keV, which corresponds
to 0.39~keV at the surface for a gravitational redshift of 0.3;
\citealt{hztb04,vdvmv04}).  If due to a proton cyclotron line, one
infers $B\simeq\expnt{6}{13}$~G \citep{hztb04}, which is substantially
larger than inferred from the spin-down.  This may simply reflect the
inadequacy of the dipole spin-down model, or the presence of higher
order multipoles.  On the other hand, based on a comparison with other
sources, \citet{vkkd+04} suggested that the absorption feature was due
to the transition from the ground state to the second excited tightly
bound state of neutral hydrogen, which would require
$B\simeq\expnt{2}{13}$~G and matches the spin-down value nicely.  If
that is the case, higher signal-to-noise spectra should reveal the
transition to the first excited state at $\sim\!0.15~$keV.

\acknowledgements We thank an anonymous referee for useful comments,
and Kaya Mori, George Pavlov, Saul Rappaport, Deepto Chakrabarty, and
Peter Woods for helpful discussions.  DLK was partially supported by a
fellowship from the Fannie and John Hertz Foundation.  We acknowledge
support through Chandra grant GO4-5082X.


\begin{thebibliography}{10}

\bibitem[{Arzoumanian} {et~al.}(1994){Arzoumanian}, {Nice}, {Taylor}, \&  {Thorsett}]{antt94}
{Arzoumanian}, Z., {Nice}, D.~J., {Taylor}, J.~H., \& {Thorsett}, S.~E. 1994,  \apj, 422, 671

\bibitem[{Becker} \& {Tr\"{u}mper}(1997){Becker} \& {Tr\"{u}mper}]{bt97}
{Becker}, W. \& {Tr\"{u}mper}, J. 1997, \aap, 326, 682

\bibitem[{Briel} \& {Pfeffermann}(1995){Briel} \& {Pfeffermann}]{bp95}
{Briel}, U.~G. \& {Pfeffermann}, E. 1995, \procspie, 2518, 120

\bibitem[{Buccheri} {et~al.}(1983){Buccheri}, {Bennett}, {Bignami}, {Bloemen},  {Boriakoff}, {Caraveo}, {Hermsen}, {Kanbach}, {Manchester}, {Masnou},  {Mayer-Hasselwander}, {Ozel}, {Paul}, {Sacco}, {Scarsi}, \&  {Strong}]{bbb+83}
{Buccheri}, R., {et al.} 1983,  \aap, 128, 245

\bibitem[{Camilo} {et~al.}(2000){Camilo}, {Kaspi}, {Lyne}, {Manchester},  {Bell}, {D'Amico}, {McKay}, \& {Crawford}]{ckl+00}
{Camilo}, F., {Kaspi}, V.~M., {Lyne}, A.~G., {Manchester}, R.~N., {Bell},  J.~F., {D'Amico}, N., {McKay}, N.~P.~F., \& {Crawford}, F. 2000, \apj, 541,  367

\bibitem[{Cox}(2000){Cox}]{allen}
{Cox}, A.~N. 2000, Allen's {A}strophysical {Q}uantities, 4th edn. (New York:  AIP Press/Springer)

\bibitem[{Cropper} {et~al.}(2004){Cropper}, {Haberl}, {Zane}, \&  {Zavlin}]{chzz04}
{Cropper}, M., {Haberl}, F., {Zane}, S., \& {Zavlin}, V.~E. 2004, \mnras, 351,  1099

\bibitem[{Cropper} {et~al.}(2001){Cropper}, {Zane}, {Ramsay}, {Haberl}, \&  {Motch}]{czr+01}
{Cropper}, M., {Zane}, S., {Ramsay}, G., {Haberl}, F., \& {Motch}, C. 2001,  \aap, 365, L302

\bibitem[Davis {et~al.}(2003)Davis, Holmes, \& Myers]{dhm03}
Davis, W., Holmes, J., \& Myers, R. 2003, in The 2003 Chandra Calibration  Workshop

\bibitem[{de Vries} {et~al.}(2004){de Vries}, {Vink}, {M{\' e}ndez}, \&  {Verbunt}]{dvvmv04}
{de Vries}, C.~P., {Vink}, J., {M{\' e}ndez}, M., \& {Verbunt}, F. 2004, \aap,  415, L31

\bibitem[{Garmire} {et~al.}(2003){Garmire}, {Bautz}, {Ford}, {Nousek}, \&  {Ricker}]{gbf+03}
{Garmire}, G.~P., {Bautz}, M.~W., {Ford}, P.~G., {Nousek}, J.~A., \& {Ricker},  G.~R. 2003, \procspie, 4851, 28

\bibitem[{Gavriil} \& {Kaspi}(2002){Gavriil} \& {Kaspi}]{gk02}
{Gavriil}, F.~P. \& {Kaspi}, V.~M. 2002, \apj, 567, 1067

\bibitem[{Gavriil} {et~al.}(2004){Gavriil}, {Kaspi}, \& {Roberts}]{gkr04}
{Gavriil}, F.~P., {Kaspi}, V.~M., \& {Roberts}, M.~S.~E. 2004, Advances in  Space Research, 33, 592

\bibitem[{Haberl}(2004){Haberl}]{haberl04}
{Haberl}, F. 2004, in XMM-Newton EPIC Consortium meeting, Palermo, 2003 October  14-16 (astro-ph/0401075)

\bibitem[{Haberl} {et~al.}(1997){Haberl}, {Motch}, {Buckley}, {Zickgraf}, \&  {Pietsch}]{hmb+97}
{Haberl}, F., {Motch}, C., {Buckley}, D. A.~H., {Zickgraf}, F.-J., \&  {Pietsch}, W. 1997, \aap, 326, 662

\bibitem[{Haberl} {et~al.}(2004){Haberl}, {Zavlin}, {Tr{\" u}mper}, \&  {Burwitz}]{hztb04}
{Haberl}, F., {Zavlin}, V.~E., {Tr{\" u}mper}, J., \& {Burwitz}, V. 2004, \aap,  419, 1077

\bibitem[{Heyl} \& {Hernquist}(1998){Heyl} \& {Hernquist}]{hh98}
{Heyl}, J.~S. \& {Hernquist}, L. 1998, \mnras, 298, L17

\bibitem[{Jansen} {et~al.}(2001){Jansen}, {Lumb}, {Altieri}, {Clavel}, {Ehle},  {Erd}, {Gabriel}, {Guainazzi}, {Gondoin}, {Much}, {Munoz}, {Santos},  {Schartel}, {Texier}, \& {Vacanti}]{jla+01}
{Jansen}, F., {et al.} 2001, \aap, 365,  L1

\bibitem[{Kaplan}(2004){Kaplan}]{kaplan04}
{Kaplan}, D.~L. 2004, Ph.D.~Thesis, California Institute of Technology

\bibitem[{Kaplan} {et~al.}(2002){Kaplan}, {Kulkarni}, {van Kerkwijk}, \&  {Marshall}]{kkvkm02}
{Kaplan}, D.~L., {Kulkarni}, S.~R., {van Kerkwijk}, M.~H., \& {Marshall}, H.~L.  2002, \apjl, 570, L79

\bibitem[{Kaplan} {et~al.}(2003){Kaplan}, {van Kerkwijk}, {Marshall},  {Jacoby}, {Kulkarni}, \& {Frail}]{kvkm+03}
{Kaplan}, D.~L., {van Kerkwijk}, M.~H., {Marshall}, H.~L., {Jacoby}, B.~A.,  {Kulkarni}, S.~R., \& {Frail}, D.~A. 2003, \apj, 590, 1008

\bibitem[{Kraft} {et~al.}(1997){Kraft}, {Chappell}, {Kenter}, {Kobayashi},  {Meehan}, {Murray}, {Zombeck}, {Fraser}, {Pearson}, {Lees}, {Brunton},  {Barbera}, {Collura}, \& {Serio}]{kck+97}
{Kraft}, R.~P., {et al.} 1997, \procspie, 3114, 53

\bibitem[{Kramer} {et~al.}(2003){Kramer}, {Lyne}, {Hobbs}, {L{\" o}hmer},  {Carr}, {Jordan}, \& {Wolszczan}]{klh+03}
{Kramer}, M., {Lyne}, A.~G., {Hobbs}, G., {L{\" o}hmer}, O., {Carr}, P.,  {Jordan}, C., \& {Wolszczan}, A. 2003, \apjl, 593, L31

\bibitem[{Lyne} {et~al.}(2000){Lyne}, {Shemar}, \& {Smith}]{lss00}
{Lyne}, A.~G., {Shemar}, S.~L., \& {Smith}, F.~G. 2000, \mnras, 315, 534

\bibitem[{Manchester} {et~al.}(2001){Manchester}, {Lyne}, {Camilo}, {Bell},  {Kaspi}, {D'Amico}, {McKay}, {Crawford}, {Stairs}, {Possenti}, {Kramer}, \&  {Sheppard}]{mlc+01}
{Manchester}, R.~N., {et al.} 2001, \mnras, 328, 17

\bibitem[{McLaughlin} {et~al.}(2003){McLaughlin}, {Stairs}, {Kaspi},  {Lorimer}, {Kramer}, {Lyne}, {Manchester}, {Camilo}, {Hobbs}, {Possenti},  {D'Amico}, \& {Faulkner}]{msk+03}
{McLaughlin}, M.~A., {et al.} 2003, \apjl, 591, L135

\bibitem[{Morris} {et~al.}(2002){Morris}, {Hobbs}, {Lyne}, {Stairs}, {Camilo},  {Manchester}, {Possenti}, {Bell}, {Kaspi}, {Amico}, {McKay}, {Crawford}, \&  {Kramer}]{mhl+02}
{Morris}, D.~J., {et al.} 2002, \mnras, 335,  275

\bibitem[{Motch} {et~al.}(2003){Motch}, {Zavlin}, \& {Haberl}]{mzh03}
{Motch}, C., {Zavlin}, V.~E., \& {Haberl}, F. 2003, \aap, 408, 323

\bibitem[{Paerels} {et~al.}(2001){Paerels}, {Mori}, {Motch}, {Haberl},  {Zavlin}, {Zane}, {Ramsay}, {Cropper}, \& {Brinkman}]{pmm+01}
{Paerels}, F., {et al.} 2001, \aap, 365, L298

\bibitem[{Pavlov} \& {Zavlin}(2003){Pavlov} \& {Zavlin}]{pz03}
{Pavlov}, G.~G. \& {Zavlin}, V.~E. 2003, in Texas in Tuscany. XXI Symposium on  Relativistic Astrophysics, ed. R.~Bandiera, R.~Maiolino, \& F.~Mannucci  (Singapore: World Scientific Publishing), 319--328 (astro-ph/0305435)

\bibitem[{Pavlov} {et~al.}(2002){Pavlov}, {Zavlin}, \& {Sanwal}]{pzs02}
{Pavlov}, G.~G., {Zavlin}, V.~E., \& {Sanwal}, D. 2002, in Neutron Stars,  Pulsars, and Supernova Remnants, ed. W.~Becker, H.~Lesch, \& J.~Tr\"{u}mper  (Garching: MPE Rep.~278), 273 (astro-ph/0206024)

\bibitem[{Str{\" u}der} {et~al.}(2001){Str{\" u}der}, {Briel}, {Dennerl},  {Hartmann}, {Kendziorra}, {Meidinger}, {Pfeffermann}, {Reppin}, {Aschenbach},  {Bornemann}, {Br{\" a}uninger}, {Burkert}, {Elender}, {Freyberg}, {Haberl},  {Hartner}, {Heuschmann}, {Hippmann}, {Kastelic}, {Kemmer}, {Kettenring},  {Kink}, {Krause}, {M{\" u}ller}, {Oppitz}, {Pietsch}, {Popp}, {Predehl},  {Read}, {Stephan}, {St{\" o}tter}, {Tr{\" u}mper}, {Holl}, {Kemmer},  {Soltau}, {St{\" o}tter}, {Weber}, {Weichert}, {von Zanthier},  {Carathanassis}, {Lutz}, {Richter}, {Solc}, {B{\" o}ttcher}, {Kuster},  {Staubert}, {Abbey}, {Holland}, {Turner}, {Balasini}, {Bignami}, {La  Palombara}, {Villa}, {Buttler}, {Gianini}, {Lain{\' e}}, {Lumb}, \&  {Dhez}]{sbd+01}
{Str{\" u}der}, L., {et al.} 2001, \aap, 365, L18

\bibitem[{Tr\"{u}mper}(1993){Tr\"{u}mper}]{trumper93}
{Tr\"{u}mper}, J. 1993, Science, 260, 1769

\bibitem[{Turner} {et~al.}(2001){Turner}, {Abbey}, {Arnaud}, {Balasini},  {Barbera}, {Belsole}, {Bennie}, {Bernard}, {Bignami}, {Boer}, {Briel},  {Butler}, {Cara}, {Chabaud}, {Cole}, {Collura}, {Conte}, {Cros}, {Denby},  {Dhez}, {Di Coco}, {Dowson}, {Ferrando}, {Ghizzardi}, {Gianotti}, {Goodall},  {Gretton}, {Griffiths}, {Hainaut}, {Hochedez}, {Holland}, {Jourdain},  {Kendziorra}, {Lagostina}, {Laine}, {La Palombara}, {Lortholary}, {Lumb},  {Marty}, {Molendi}, {Pigot}, {Poindron}, {Pounds}, {Reeves}, {Reppin},  {Rothenflug}, {Salvetat}, {Sauvageot}, {Schmitt}, {Sembay}, {Short},  {Spragg}, {Stephen}, {Str{\" u}der}, {Tiengo}, {Trifoglio}, {Tr{\" u}mper},  {Vercellone}, {Vigroux}, {Villa}, {Ward}, {Whitehead}, \& {Zonca}]{taa+01}
{Turner}, M.~J.~L., {et al.} 2001, \aap, 365, L27

\bibitem[{van Kerkwijk} {et~al.}(2004){van Kerkwijk}, {Kaplan}, {Durant},  {Kulkarni}, \& {Paerels}]{vkkd+04}
{van Kerkwijk}, M.~H., {Kaplan}, D.~L., {Durant}, M., {Kulkarni}, S.~R., \&  {Paerels}, F. 2004, \apj, 608, 432

\bibitem[{van Riper} {et~al.}(1991){van Riper}, {Epstein}, \&  {Miller}]{vrem91}
{van Riper}, K.~A., {Epstein}, R.~I., \& {Miller}, G.~S. 1991, \apjl, 381, L47

\bibitem[{Vink} {et~al.}(2004){Vink}, {de Vries}, {M{\' e}ndez}, \&  {Verbunt}]{vdvmv04}
{Vink}, J., {de Vries}, C.~P., {M{\' e}ndez}, M., \& {Verbunt}, F. 2004, \apjl,  609, L75

\bibitem[{Vlemmings} {et~al.}(2004){Vlemmings}, {Cordes}, \&  {Chatterjee}]{vcc04}
{Vlemmings}, W.~H.~T., {Cordes}, J.~M., \& {Chatterjee}, S. 2004, \apj, 610,  402

\bibitem[{Voges} {et~al.}(1996){Voges}, {Aschenbach}, {Boller}, {Brauninger},  {Briel}, {Burkert}, {Dennerl}, {Englhauser}, {Gruber}, {Haberl}, {Hartner},  {Hasinger}, {Kurster}, {Pfeffermann}, {Pietsch}, {Predehl}, {Rosso},  {Schmitt}, {Tr\"{u}mper}, \& {Zimmermann}]{rbs}
{Voges}, W., {et al.} 1996, \iaucirc, 6420, 2

\bibitem[{Vranesevic} {et~al.}(2004){Vranesevic}, {Manchester}, {Lorimer},  {Hobbs}, {Lyne}, {Kramer}, {Camilo}, {Stairs}, {Kaspi}, {D'Amico},  {Possenti}, {Crawford}, {Faulkner}, \& {McLaughlin}]{vml+04}
{Vranesevic}, N., {et al.} 2004, \apjl, 617, L139

\bibitem[{Weisskopf} {et~al.}(2000){Weisskopf}, {Tananbaum}, {Van Speybroeck},  \& {O'Dell}]{wtvso00}
{Weisskopf}, M.~C., {Tananbaum}, H.~D., {Van Speybroeck}, L.~P., \& {O'Dell},  S.~L. 2000, \procspie, 4012, 2

\bibitem[{Woods} {et~al.}(2000){Woods}, {Kouveliotou}, {Finger},  {G\"{o}\u{g}\"{u}\c{s}}, {Scott}, {Dieters}, {Thompson}, {Duncan}, {Hurley},  {Strohmayer}, {Swank}, \& {Murakami}]{wkf+00}
{Woods}, P.~M., {et al.} 2000, \apjl, 535, L55

\bibitem[{Young} {et~al.}(1999){Young}, {Manchester}, \& {Johnston}]{ymj99}
{Young}, M.~D., {Manchester}, R.~N., \& {Johnston}, S. 1999, \nat,
400, 848

\bibitem[{Zane} {et~al.}(2002){Zane}, {Haberl}, {Cropper}, {Zavlin}, {Lumb},  {Sembay}, \& {Motch}]{zhc+02}
{Zane}, S., {Haberl}, F., {Cropper}, M., {Zavlin}, V.~E., {Lumb}, D., {Sembay},  S., \& {Motch}, C. 2002, \mnras, 334, 345

\bibitem[{Zavlin} {et~al.}(2000){Zavlin}, {Pavlov}, {Sanwal}, \&  {Tr{\"u}mper}]{zpst00}
{Zavlin}, V.~E., {Pavlov}, G.~G., {Sanwal}, D., \& {Tr{\"u}mper}, J. 2000,  \apjl, 540, L25

\bibitem[{Zombeck} {et~al.}(1995){Zombeck}, {David}, {Harnden}, \&  {Kearns}]{zdhk95}
{Zombeck}, M.~V., {David}, L.~P., {Harnden}, F.~R., \& {Kearns}, K. 1995,  \procspie, 2518, 304

\end{thebibliography}

\end{document}